\documentclass[12pt]{article}
\usepackage{latexsym,amsmath,amssymb,amsfonts,amsthm,bbm, hyperref, blkarray}
\usepackage{natbib}
\usepackage{enumerate}
\usepackage{graphicx}
\usepackage{graphics,enumerate}
\usepackage[x11names,svgnames]{xcolor}

\usepackage{amsmath, amsfonts, capt-of, algorithm, algpseudocode, tikz, color}

\begin{document}

\title{Estimating whole brain dynamics using spectral clustering}
\author{Ivor Cribben and Yi Yu\\
Alberta School of Business, Canada\\
Statistical Laboratory, University of Cambridge, U.K.}

\maketitle
\begin{abstract}
 
The estimation of time-varying networks for functional Magnetic Resonance Imaging (fMRI) data sets is of increasing importance and interest.  In this work, we formulate the problem in a high-dimensional time series framework and introduce a data-driven method, namely \emph{Network Change Points Detection (NCPD)}, which detects change points in the network structure of a multivariate time series, with each component of the time series represented by a node in the network.  NCPD is applied to various simulated data and a resting-state fMRI data set.  This new methodology also allows us to identify common functional states within and across subjects.  Finally, NCPD promises to offer a deep insight into the large-scale characterisations and dynamics of the brain. 

Keywords: Spectral clustering; Change point analysis; Network change points; Stationary bootstrap; fMRI; Resting-state data.

\end{abstract}

\section{Introduction}\setcounter{equation}{0}\label{sec:int} \setcounter{figure}{0}

In the `Big Data' era, time-varying network models are used to solve many important problems.  In particular, a great current challenge in neuroscience is the reconstruction of the dynamic manner in which brain regions interact with one another in both task-based and resting-state functional magnetic resonance imaging (fMRI) studies. This reconstruction has the ability to have a major impact on the understanding of the functional organisation of the brain.  fMRI is a neuroimaging technique that indirectly measures brain activity.  \cite{Ogawa} introduced the blood-oxygen-level dependent (BOLD) contrast, which is currently the primary form of fMRI due to its high spatial resolution and its non-invasiveness. BOLD is based on the dependence between blood flow in the brain and neuronal activation.  In other words, when a certain brain region is active, extra blood flows to this region.  In particular, the brain is usually parcellated into small cubic regions roughly a few millimetres in size called voxels and the brain activity is measured in each voxel across a sequence of scans with time series as outputs. 
 
Functional connectivity (FC) analysis examines functional associations between time series pairs in specified voxels or regions \citep{biswal}. The simplest methods for estimating FC include the use of covariance, correlation and precision matrices \citep{cummine}.  FC can also be estimated using spectral measures such as coherence and partial coherence \citep{cribbenfiecas,fiecas}.  In addition, the FC between brain voxels or regions can be represented by an interconnected brain network. Here, vertices and edges represent brain region time series and their FC, respectively. The idea of studying the brain as an FC network is helpful as it can be viewed as a system with various interacting regions which produce complex behaviours. Similar to other biological networks, understanding the complex network organisation of the brain can lead to profound clinical implications \citep{sporns,bullmore}.

All of these methods, however, assume that the data are stationary over time, that is, the dependence or the FC between brain regions remains constant throughout the experiment. Although this assumption is convenient for both statistical estimation and computational reasons, it presents a simplified version of a highly integrated and dynamic phenomenon.  Evidence of the non-stationary behaviour of time series from brain activity has been observed not only in task-based fMRI experiments \citep{cribben2,cribben3,debener,fox,eichele,sadaghiani}, but also prominently in resting-state data \citep{delamillieure,doucet}. 

This evidence has led to an increase in the number of statistical methods that estimate the time-varying or dynamic connectivity.  The covariance, correlation and precision matrix approaches discussed above all have a natural time-varying analogue.  Using a moving-window, these approaches begin at the first time point, then a block of a fixed number of time points are selected and all the data points within the block are used to estimate the FC.  The window is then shifted a certain number of time points and the FC is estimated on the new data set. By shifting the window to the end of the experimental time course, researchers can estimate the time-varying FC.  Many research papers have considered this approach.   \cite{chang}, \cite{kiviniemi} and \cite{hutchison2} investigated the non-stationary behaviour of resting-state connectivity using a moving-window approach, based on a time-frequency coherence analysis with wavelet transforms, an independent component analysis and a correlation analysis, respectively.   \cite{leonardi} studied whole brain dynamic FC using a moving-window and a principal component analysis technique that is applied to resting-state data.  \cite{leonardi2} introduced a data-driven multivariate method, namely higher-order singular value decomposition, which models whole brain networks from group-level time-varying FC data using a moving-window based on a tensor decomposition.  \cite{allen}, \cite{handwerker}, \cite{jones} and \cite{sakouglu} considered a group independent component analysis \citep{calhoun} to decompose multi-subject resting-state data into functional brain regions, and a moving-window and $k$-means clustering of the windowed correlation matrices to study whole brain time-varying networks. 

While the moving-window approach can be used to observe time-varying FC, and is computationally feasible, it also has limitations \citep{hutchison}.  For example, the choice of block size is crucial and sensitive, as different block sizes can lead to different FC patterns.  Another pitfall is that the technique gives equal weight to all $k$ neighbouring time points and 0 weight to all the others \citep{lindquist:yuting}.  In order to estimate time-varying FC without the use of sliding windows, \cite{zhang} proposed the dynamic Bayesian variable partition model that estimates and models multivariate dynamic functional interactions using a unified Bayesian framework.  This method first detects the temporal boundaries of piecewise quasi-stable functional interaction patterns, which are then modelled by representative signature patterns and whose temporal transitions are characterised by finite-state transition machines.  

There are a number of methods that utilize change point procedures for estimating the time-varying connectivity between brain imaging signals, including Dynamic Connectivity Regression (DCR:~\citealp{cribben1, cribben2}), FreSpeD \citep{schroder}, and a novel statistical method for detecting change points in multivariate time series \citep{kirch}.  \cite{schroder} employ a multivariate cumulative sum (CUSUM)-type procedure to detect change points in autospectra and coherences for multivariate time series.  Their methods allows for the segmentation of the multivariate time series but also for the direct interpretation of the change in the sense that the change point can be assigned to one or multiple time series (or Electroencephalogram channels) and frequency bands.   \cite{kirch} consider the At Most One Change (AMOC) setting and the epidemic setting (two change points, where the process reverts back to the original regime after the second change point) and provide some theoretical results.

All of these techniques are based on different methodologies but each of them performs very well in their own right.  However, they have limitations.   The most obvious is that they are all restricted by the number of time series from either the channels or brain regions.  For DCR, the algorithm begins to slow after 50 time series.  In addition, FreSpeD considers only 21 time series channels in its application and, like the DCR method, it is limited by the minimum separability assumption, which means that there has to be a certain distance between change points.  Finally,  the method of \cite{kirch} is also restricted by the number of time series they can include because the proposed test statistics require the estimation of the inverse of the long run auto-covariance, which is particularly difficult in higher-dimensional settings and even more problematic in the multivariate case because of the number of entries in the positive-definite weight matrix.  Their method also focuses on changes in the model parameters, which is limiting as it is difficult to interpret a change in a parameter. 

In this paper, our aim is not only to detect the network structural changes along the experimental time course, but also to represent the high-dimensional brain imaging data in a low dimensional clustering structure; in other words, we are interested in combining the research areas of change point detection in time series analysis and community detection in network analysis.  Recently, both change point detection in time series and community detection in network statistics have become topical areas (see e.g. \citealp{FrickEtal2014}, \citealp{ChoFryzlewicz2015}, \citealp{WangSamworth2016}, for some up-to-date work on change point detection, and e.g. \citealp{Newman2006}, \citealp{WangBickel2015}, \citealp{Jin2015} for some recent work on community detection).  The essence of these two areas is to partition the data set into different clusters/segments that share some fundamental similarities but differ from the other clusters/segments.  Specifically, change point detection is the segmentation of non-stationary time series into several stationary segments while community detection is the partitioning of complex networks into several tightly-knit clusters.  

To this end, we introduce a data-driven method, namely \emph{network change point detection (NCPD)}, the detailed methodology and algorithm of which are explained in Section~\ref{sec:modelal}.  NCPD's strength, unlike the other change point methods above, is that it can consider thousands of time series and in particular the case where the number of brain regions well exceeds the number of time points in the experiment, i.e., $p \gg T$, where $p$ is the number of regions of interest or voxels and $T$ is the number of time points. Using NCPD, one can, therefore, consider whole brain dynamics, which departs from the moving window technique and promises to offer deeper insight into the large scale characterisations of the whole brain. 

We apply the new method to a resting-state fMRI experiment.  Dynamic FC is prominent in the resting-state when mental activity is unconstrained.  This analysis has led to the robust identification of cognitive states at rest.   NCPD not only allows for the estimation of time-varying connectivity but also finds common cognitive states that recur in time and across subjects in a group study.  By unveiling the time-varying cognitive states of both controls and subjects with neuropsychiatric diseases such as Alzheimer's, dementia, autism and schizophrenia using NCPD, we can compare their FC patterns and endeavour to develop new understandings of these diseases.  NCPD is, to the best of our knowledge, the first paper to consider estimating change points for time evolving community network structure in a multivariate time series context.  Although this paper is inspired by and developed for brain connectivity studies, our proposed method pertains to a general setting and can also be used in a variety of situations where one wishes to study the evolution of a high dimensional network over time.  

The rest of this paper is organised as follows.  The required notation for this paper is introduced in Section~\ref{sec:notation}.  NCPD is outlined in Section~\ref{sec:modelal}, with simulations and real data analysis presented in Sections~\ref{sec:sims} and \ref{sec:rsfmridata}, respectively.  We conclude this paper with a discussion in Section~\ref{sec:discussion}.

\subsection{Notation}\label{sec:notation}

In this subsection, we introduce the standard graph-theoretic notation.  We do not distinguish between the terms `network' and `graph' in this paper.  Let $G := (V,E)$ denote a $p$-node undirected simple graph, where $V := \{1, \ldots, p\}$ and $E := \{(i, j),\, 1\leq i< j\leq p\}$ are the collections of vertices and edges, respectively.  A $K$-\emph{partition} is a pairwise disjoint collection $\{V_k: k =1,\ldots,K\}$ of non-empty subsets of $V$ such that $V = V_1 \sqcup \ldots \sqcup V_K$, where $\sqcup$ denotes disjoint union.

The \emph{adjacency matrix} of $G$ is denoted by $A:= (A_{ij})_{1\leq i,j \leq p}$, where $A_{ij} = 1$ if $(i, j) \in E$ or $(j, i) \in E$, otherwise $A_{ij} = 0$. The \emph{degree} of vertex $i\in V$ is $d_i := \sum_{j=1}^p A_{ij}$, and the degree matrix is $D := \mathrm{diag}(d_1,\ldots,d_p)$. The vital tool in spectral clustering is the \emph{Laplacian matrix} \citep{Chung1997}, which is defined as 
\begin{equation}\label{eq-laplacian}
L := D - A,
\end{equation}
with eigenvalues $0 = \lambda_1 \leq \ldots \leq \lambda_p$ and corresponding unit-length eigenvectors $\boldsymbol{v}_1, \ldots, \boldsymbol{v}_p$. 

In the rest of this paper, we denote true covariance matrices and sample correlation matrices by $\Sigma$ and $R$ respectively.  For a matrix $M$, denote $M_{(i)}$ as the $i$th row and $M_{(a:b)}$ as the sub-matrix of $M$ consisting of the $a$th to $b$th rows of $M$, where $a < b$.

\section{Methodology} \label{sec:modelal}

In this section, we introduce the NCPD method.  Motivated by, but not restricted to, the brain dynamics analysis, we use `nodes' instead of `voxels' or `regions'.  To start this section, we illustrate the algorithm and then elaborate more on the details.  The input of the following algorithm includes:
\begin{itemize}
\item a data matrix $Y \in \mathbb{R}^{T\times p}$, where $T$ and $p$ are the numbers of time points and nodes respectively;
\item the pre-specified number of communities $K$;
\item a collection of candidate change points $\boldsymbol{\delta} := \{\delta_1, \ldots, \delta_m\} \subset \{1, \ldots, T\}$;
\item a pre-specified significance thresholding $\alpha \in (0, 1)$.
\end{itemize}

\begin{algorithm}
\caption{Network Change Point Detection}\label{Alg:main}
 \begin{algorithmic}[1]
 \Procedure{NCPD}{$Y, K, \boldsymbol{\delta}, \alpha$}
 \For{$j = 1, \ldots, m$}
 \State $(Y_{\mathrm{L}},  Y_{\mathrm{R}}) \gets$ $(Y_{(1: \delta_j)}, Y_{(\delta_j + 1 : T)})$
 \State $(R_{\mathrm{L}}, R_{\mathrm{R}}) \gets (\mathrm{corr}(Y_{\mathrm{L}}), \mathrm{corr}(Y_{\mathrm{R}}))$
 \State $(\boldsymbol{z}_{\mathrm{L}}, C_{\mathrm{L}}) \gets \mathrm{SpectralClustering}(R_{\mathrm{L}}, K)$; $(\boldsymbol{z}_{\mathrm{R}}, C_{\mathrm{R}}) \gets \mathrm{SpectralClustering}(R_{\mathrm{R}}, K)$
     \For{$i = 1, \ldots, p$}
     \State $(U_{\mathrm{L}}^i, U_{\mathrm{R}}^i) \gets (C_{\mathrm{L}}^{\boldsymbol{z}_{\mathrm{L}},i}, C_{\mathrm{R}}^{\boldsymbol{z}_{\mathrm{R}},i})$
     \EndFor
 \State $\gamma_j \gets$ sum of the singular values of $U_{\mathrm{L}}^{\top}U_{\mathrm{R}}$   
 \State $\mathrm{FLAG}_j \gets \mathrm{StationaryBootstrap}(\alpha, \gamma_j)$
 \EndFor
 \State \textbf{return} $\{\delta_j: j\in\{1, \ldots, m\}, \mathrm{FLAG}_j = \mathrm{significant}\}$
 \EndProcedure
 \end{algorithmic}
\end{algorithm}

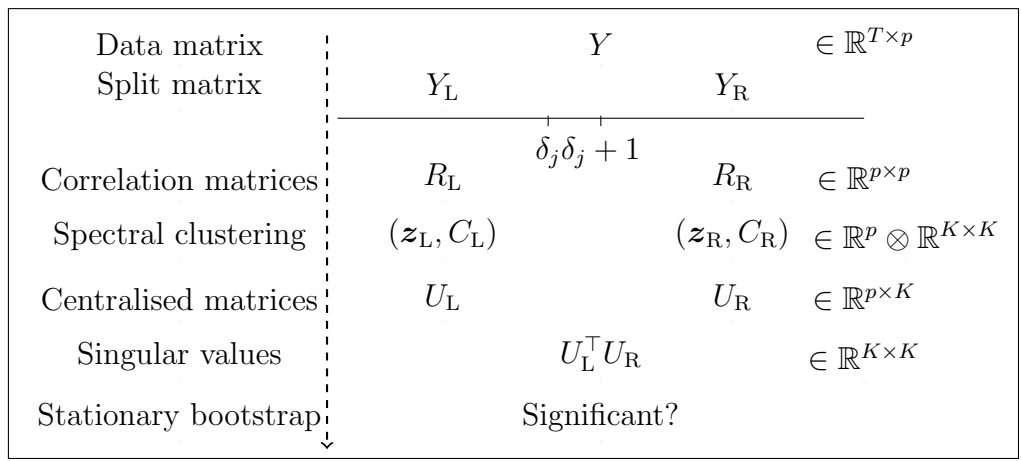
\begin{figure}[h]
\centering
\fbox{
\begin{tikzpicture}[y=.2cm, x=.7cm]
	\draw (0,0) -- (10,0);
    	\draw (5,3.5) -- (5,3.51)
			node[anchor=south] {$Y$};    	
	\draw (10,3.5) -- (10,3.51)
			node[anchor=south] {$\in \mathbb{R}^{T \times p}$};
	\draw (-3, 3.5) -- (-3, 3.51)
			node[anchor=south] {Data matrix};	
    	\draw (4,1pt) -- (4,-3pt)
			node[anchor=north] {$\delta_j$};
    	\draw (5,1pt) -- (5,-3pt)
			node[anchor=north] {$\delta_j + 1$};		\draw (2,0.5) -- (2,0.51)
			node[anchor=south] {$Y_{\mathrm{L}}$};	\draw (7.5,0.5) -- (7.5,0.51)
			node[anchor=south] {$Y_{\mathrm{R}}$};	\draw (-3, 0.5) -- (-3, 0.51)
			node[anchor=south] {Split matrix};
	\draw (-3, -5.5) -- (-3, -5.51)
			node[anchor=south] {Correlation matrices};
	\draw (2,-5.5) -- (2,-5.51)
			node[anchor=south] {$R_{\mathrm{L}}$};	\draw (7.5,-5.5) -- (7.5,-5.51)
			node[anchor=south] {$R_{\mathrm{R}}$};
	\draw (10,-5.5) -- (10,-5.51)
			node[anchor=south] {$\in \mathbb{R}^{p \times p}$};	\draw (-3, -9.5) -- (-3, -9.51)
			node[anchor=south] {Spectral clustering};
	\draw (2,-9.5) -- (2,-9.51)
			node[anchor=south] {$(\boldsymbol{z}_{\mathrm{L}}, C_{\mathrm{L}})$};	
	\draw (7.5,-9.5) -- (7.5,-9.51)
			node[anchor=south] {$(\boldsymbol{z}_{\mathrm{R}}, C_{\mathrm{R}})$};
	\draw (10.8,-9.5) -- (10.8,-9.51)
			node[anchor=south] {$\in \mathbb{R}^p \otimes \mathbb{R}^{K\times K}$};	\draw (-3, -13.5) -- (-3, -13.51)
			node[anchor=south] {Centralised matrices};
	\draw (2,-13.5) -- (2,-13.51)
			node[anchor=south] {$U_{\mathrm{L}}$};	\draw (7.5,-13.5) -- (7.5,-13.51)
			node[anchor=south] {$U_{\mathrm{R}}$};
	\draw (10,-13.5) -- (10,-13.51)
			node[anchor=south] {$\in \mathbb{R}^{p \times K}$};	\draw (-3, -17.5) -- (-3, -17.51)
			node[anchor=south] {Singular values};
	\draw (5, -17.5) -- (5, -17.51)
			node[anchor=south] {$U_{\mathrm{L}}^{\top}U_{\mathrm{R}}$};	\draw (10,-17.5) -- (10,-17.51)
			node[anchor=south] {$\in \mathbb{R}^{K \times K}$};	\draw (-3, -21.5) -- (-3, -21.51)
			node[anchor=south] {Stationary bootstrap};	\draw (5, -21.5) -- (5, -21.51)
			node[anchor=south] {Significant?};
    \draw[dashed, thick,->] (-0.2, 5.5) -- (-0.2, -22);				
\end{tikzpicture}}
\caption{The work flow of the NCPD algorithm}\label{Fig:work-flow}
\end{figure}

The work-flow and the pseudo-code of NCPD are given in Figure~\ref{Fig:work-flow} and Algorithm~\ref{Alg:main}, respectively.

\subsection{Spectral clustering}

Spectral clustering \citep{DonathHoffman1973} is a computationally feasible and nonparametric method widely used in community detection in network statistics. In an undirected simple network $G = (V, E)$, we believe that the vertices are tightly-knit within the communities and loosely-connected between communities. A natural criterion in the recovery of the community structure is to minimise the number of between community edges with the sizes of the communities as the normalisation, namely the ratio cut \citep{WeiCheng1991}, which is defined as
\[
\mathrm{RCut}(V_1,\ldots,V_K) := \sum_{k = 1}^K \frac{W(V_k, V_k^c)}{|V_k|},
\]
where $W(V_k,V_k^c) := \sum_{i \in V_k, j \in V_k^c} A_{ij}$ is the total number of edges connecting $V_k$ and its complement $V_k^c$. However, seeking the partition minimising $\mathrm{RCut}$ is an NP-hard problem \citep{GJS1976}, while the spectral clustering is its convex relaxation \citep{VonLuxburg}.

In a high-dimensional time series data set, $Y\in\mathbb{R}^{T\times p}$, with each component (each column of $Y$) a node in a collaboration network, the connectivity network in the given time period is therefore captured by its correlation matrix $R \in \mathbb{R}^{p \times p}$. Treating the correlation matrix $R$ as the adjacency matrix, its corresponding Laplacian matrix $L$ can be computed following Equation~\eqref{eq-laplacian}. 
 
Spectral clustering unveils the community structure by exploiting the eigen-structure of the Laplacian matrix $L$.  Let $V$ consist of the unit-length eigenvectors that are associated with the $K$ smallest eigenvalues of $L$, namely $V = (\boldsymbol{v}_1, \ldots, \boldsymbol{v}_K)$, which is a $K$-dimensional embedding of the $p$-dimensional network. The information of each node is therefore captured by a point in $\mathbb{R}^K$. To discover the community structure, $k$-means clustering is applied to the rows of $V$ and returns the community labels $\boldsymbol{z} := (z_1, \ldots, z_p) \in \{1, \ldots, K\}^p$ and $K$ centroids.  A generic spectral clustering algorithm is provided in Algorithm~\ref{Alg:SC} with the input being the adjacency matrix $A$ and the pre-specified community number $K$, and the outputs are the estimated labels $\boldsymbol{z}$ and centroids of the communities $C$.

\begin{algorithm}
\caption{Generic Spectral Clustering}\label{Alg:SC}
 \begin{algorithmic}[1]
 \Procedure{SpectralClustering}{$A \in \mathbb{R}^{p \times p}$, $K$}
 \State $d_i \gets \sum_{j = 1}^p A_{ij}$
 \State $D \gets \mathrm{diag}\{d_1, \ldots, d_p\}$
 \State $L \gets D - A$
 \State $\{\boldsymbol{v}_1, \ldots, \boldsymbol{v}_K\} \gets$ unit-length eigenvectors of $L$ which are associated with the $K$ smallest eigenvalues of $L$
 \State $V \gets (\boldsymbol{v}_1, \ldots, \boldsymbol{v}_K)$
 \State cluster labels for all nodes and centroids of $K$ communities $(\boldsymbol{z}, C) \in \mathbb{R}^p \otimes \mathbb{R}^{K\times K} \gets$ results of $k$-means clustering on the rows of $V$ with $K$ centres
 \State \textbf{return} $(\boldsymbol{z}, C)$
 \EndProcedure
 \end{algorithmic}
\end{algorithm}

Note that, in high-dimensional data analysis, penalised precision matrices are often used to study the underlying graphs, when the assumption is only a few pairs out of $p(p-1)/2$ pairs are correlated, for a $p$-node network.  However, in this paper, we propose that sparsity means that a low dimension formation appears in the community structure.  A $p$-node and $K$-community network can have related pairs of order $O(p^2)$, which is not achievable by penalising precision matrices.

\subsection{Singular values}

Recall that the main purpose of this paper is to detect change points in terms of the network structure. Spectral clustering unveils the community structure and reduces the dimension of the data sets from $p$ -- the number of nodes -- to $K$ -- the number of communities. The next task is to evaluate the deviance between the network before and after a certain candidate change point. 

A natural measurement of the difference between two spaces spanned by the columns of two matrices respectively is the \emph{principal angles}: if $V, W \in \mathbb{R}^{p \times K}$ both consist of orthonormal columns, then the $K$ principal angles between their column spaces are $\cos^{-1} \sigma_1,\ldots,\cos^{-1}\sigma_K$, where $\sigma_1 \geq \cdots \geq \sigma_K$ are the singular values of $V^\top W$.  Principal angles between pairs of subspaces can be regarded as natural generalisations of acute angles between pairs of vectors. 

The rationale behind principal angles in community detection is the community label invariance. Since the columns of matrix $U$ represent the communities, the measurement should be invariant in terms of the rotation of the columns, i.e. right multiplied by any orthogonal matrix $O \in \mathbb{R}^{K \times K}$ . For any orthogonal matrix $O \in \mathbb{R}^{K \times K}$, matrices $V^\top W$ and $V^\top WO$ have the same singular values. 

In our problem, we are interested in network structure changes. Spectral clustering on the Laplacian matrix provides the community information. For each candidate change point, we then construct new matrices $U_{\mathrm{L}}$ and $U_{\mathrm{R}}$, whose rows are the corresponding centroids. The column spaces of $U_{\mathrm{L}}$ and $U_{\mathrm{R}}$ encode the averaged location information, so we do not impose the condition that the columns have to be orthonormal. However, for a certain change point candidate, the singular values of $U_{\mathrm{L}}^{\top}U_{\mathrm{R}}$ still unveil the deviance in terms of the network structure between the two networks separated by the candidate change point.   We denote by $\gamma = \gamma(U_{\mathrm{L}}, U_{\mathrm{R}}) := \sum_{k = 1}^K \sigma_k$, where $\{\sigma_1, \ldots, \sigma_K\}$ are the singular values of $U_{\mathrm{L}}^{\top}U_{\mathrm{R}}$.  In the sequel, the subscript of $\gamma$ indicates the corresponding candidate change point.

Since the singular values are the cosine values of the principal angles, the smaller $\gamma$ is, the more prominent the difference between the two subspaces is; therefore, a change point is expected to have the smallest $\gamma$ value. 

\subsection{Selection and stopping criteria}\label{sec:selec-cri}

In principal, we treat every time point as a candidate change point and compare the deviance of the networks before and after it. However, in practice, we need enough data points to construct a network. Denote $n_{\min}$ as the lower limit of the number of time points needed to construct a network and recall that $T$ is the total number of time points available, we calculate the $\gamma$ criterion values associated with all the points from $n_{\min}$ to $T - n_{\min}$ and pick the time point with the smallest $\gamma$ after deleting the \emph{outliers}, the definition of which we will now specify. 

\begin{figure}[h]
\centering
\makebox{\includegraphics[width = 0.8\textwidth]{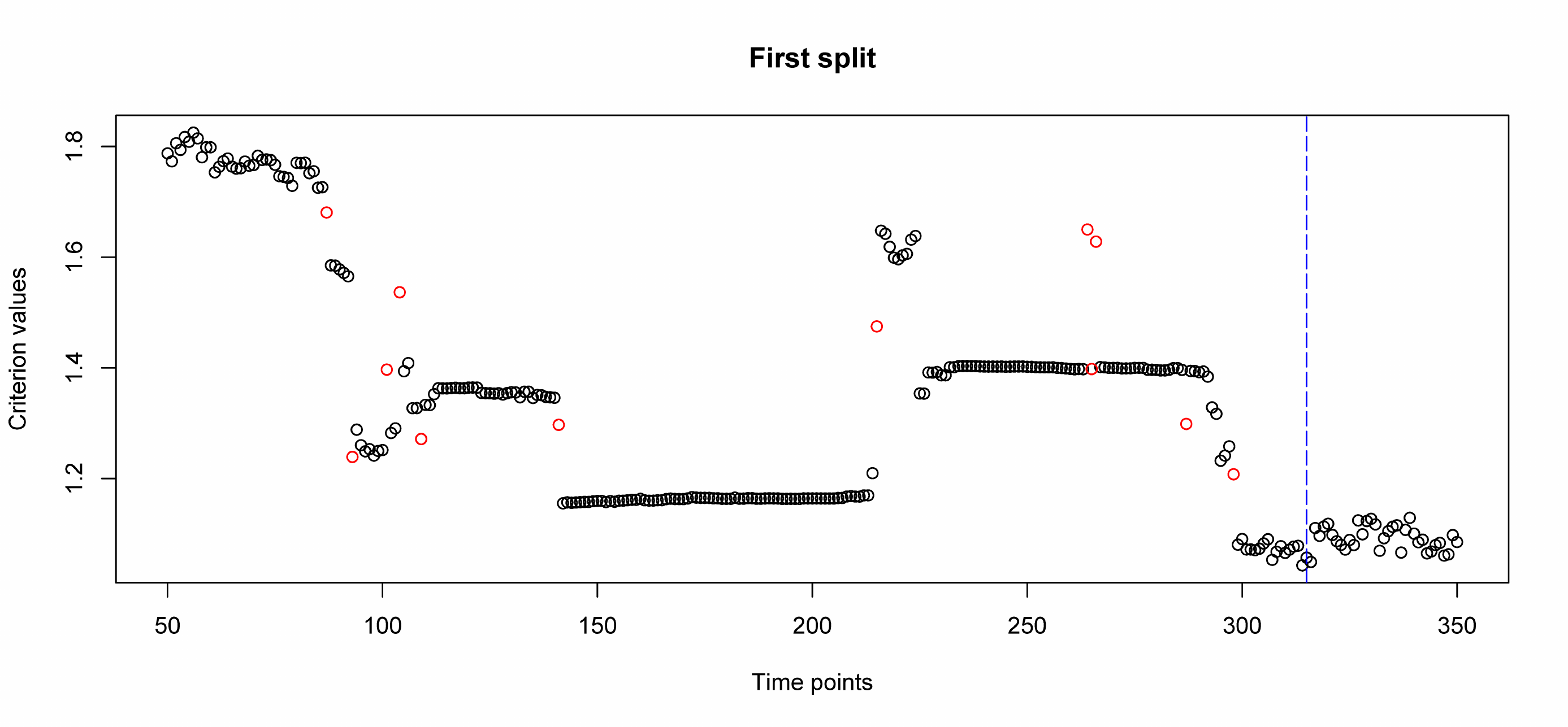}}
\makebox{\includegraphics[width = 0.8\textwidth]{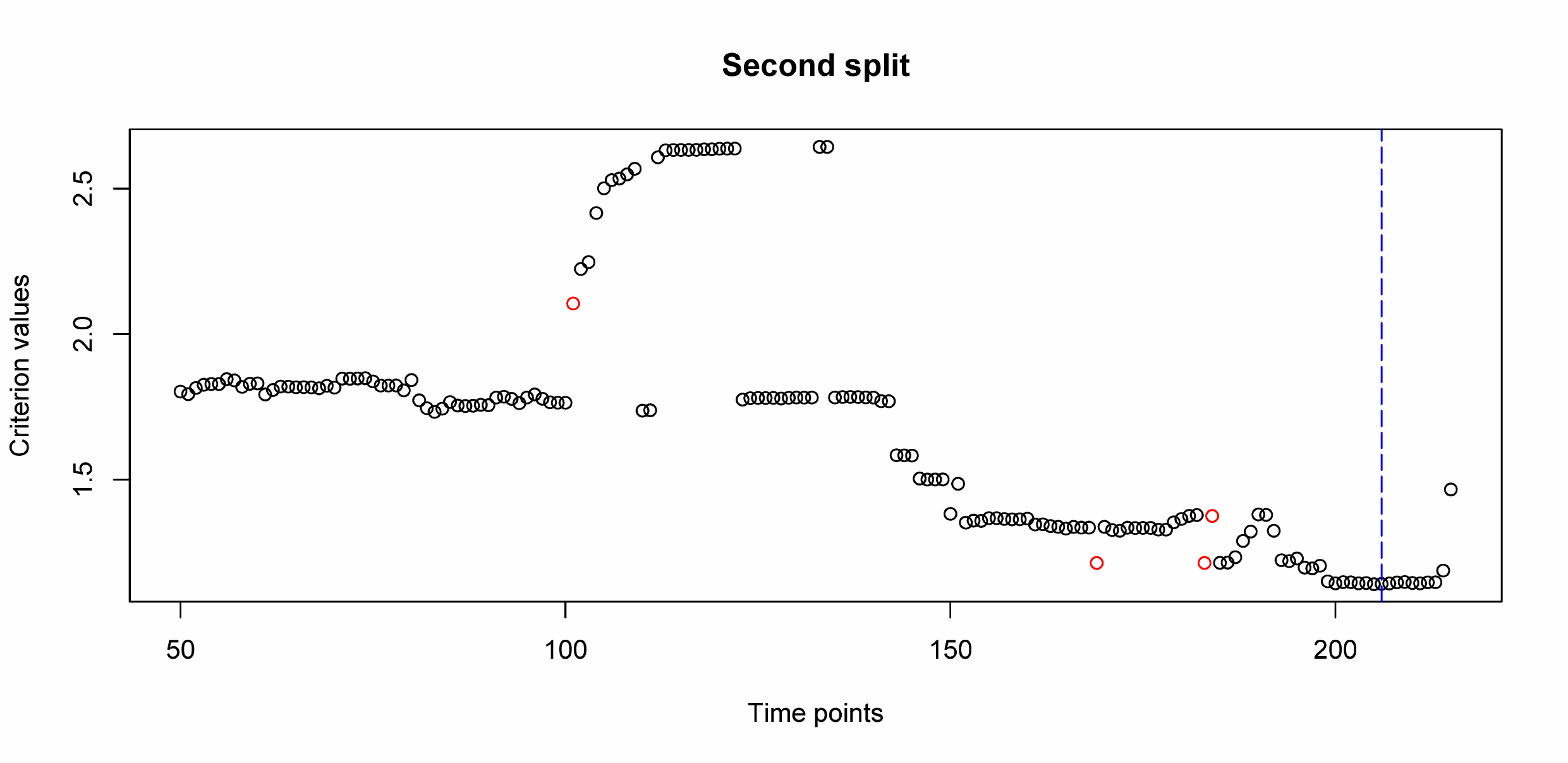}}
\caption{\label{fig:cri-value} Criterion values for each candidate change point.  Red circles represent the outliers and the blue dashed lines represent the estimated change points locations.}
\end{figure}

We use an example to illustrate the necessity of the exclusion of the outliers.  In Figure~\ref{fig:cri-value}, the $\gamma$ criterion values of all candidate change points are presented. The upper and lower panels represent the first and second change points, respectively.  The true change points occur at time points 300 and 200 respectively in these two plots.  In addition, ideally the change along the time-axis should be smooth. In practice, as we can see in Figure~\ref{fig:cri-value}, there are some points that have very different values from those of their neighbours, i.e. those coloured in red. 

For candidate change points $j$, $2\leq j \leq m-1$, define the outlier detection value $\eta_j := \max\{|\gamma_j - \gamma_{j-1}|, |\gamma_j - \gamma_{j+1}|\}$, $\eta_1 := |\gamma_2 - \gamma_1|$ and $\eta_m := |\gamma_m - \gamma_{m-1}|$. The outliers are those points that have extremely large values of $\eta$, i.e. those which are associated with the largest 5\% of the $\eta$ values and are denoted by red points in Figure~\ref{fig:cri-value}.  

\begin{figure}[ht]
\centering
\makebox{\includegraphics[scale = 0.6]{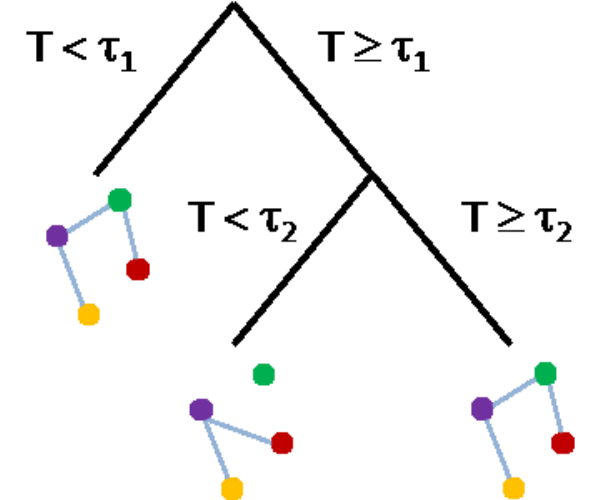}}
\caption{\label{fig:split} An illustration of the exhaustive search and split algorithm.}
\end{figure}

We run the algorithm exhaustively until the available time points are fewer than the pre-specified threshold, and construct networks for each segment; an illustration is given in Figure~\ref{fig:split}.

\subsection{Inference on change points}\label{sec:inference}
In this subsection we discuss an inferential procedure for the cosine of the principal angles between the two subspaces at the candidate change points.  As the candidate change points are found,  we estimate confidence bounds for the $\gamma$ criterion values at every candidate change point using the stationary bootstrap \citep{politis}.  An assumption of the proposed methodology is the presence of autocorrelation in the individual time series of the data matrix $Y$.  Hence, by using the stationary bootstrap or resampling blocks of data points, the dependence structure inherent in the data remains intact and the correct confidence bounds are calculated.  The stationary bootstrap is an adaptation of the block bootstrap \citep{Liu:Singh} but it resamples blocks of data that are of varying block sizes.  

The stationary bootstrap procedures aim to detect whether the smallest criterion value $\gamma$, over the time period being studied after outlier deletion, is significant.  Without loss of generality, we assume the time period being studied is from $1$ to $T$, and the first change point occurs at time point $\delta$, which has the smallest criterion value $\gamma$ after outlier deletions.  The procedure then generates pseudo-samples and conducts statistical inference based on these.  To describe the algorithm, we adopt the method in \cite{politis}, letting $B_{i, b} := \{Y_{(i)}, Y_{(i+1)}, \ldots, Y_{(i+b-1)}\}$ be the block consisting of $b$ consecutive time points starting from the $i$th one.   In the case $j > T$, set $Y_{(j)} = Y_{(i)}$ with $i := j(\mathrm{mod} T)$ and $Y_{(0)} := Y_{(T)}$.  A pseudo-sample $\bigl\{Y_{(1)}^{*}, Y_{(2)}^{*}, \ldots, Y_{(T)}^{*}\bigr\}$ is generated as follows:
\begin{enumerate}
\item independently generate $M$ realisations $L_1, \ldots, L_M$ from the geometric distribution with parameter $p \in (0, 1)$,  such that $\sum_{i= 1}^{M-1}L_i < T$ and $\sum_{i=1}^M L_i \geq T$;
\item independently generate $M$ realisations $I_1, \ldots, I_M$ from the discrete uniform distribution on $\{1, \ldots, T\}$;
\item the pseudo-sample is the first chunk of $T$ realisations in $\{B_{I_1, L_1}, \ldots, B_{I_M, L_M}\}$.
\end{enumerate}

To test whether the  $\gamma$ criterion value at the candidate change point $\delta$ is significant, we generate many, say 1000, pseudo-samples $Y_{(1:T)}^{*}$, for each of which, a new $\gamma_{\delta}$ -- the criterion value at $\delta$ -- is calculated.  The null hypothesis is that the time point $\delta$ is not a change point; therefore for a pre-specified $p$-value $\alpha \in (0, 1)$ ($\alpha = 0.05$ in the numerical studies in this paper), we calculate $c_\alpha$ as the $100\alpha$th empirical quantile of the stationary bootstrap distribution of $\gamma_\delta$.  If the observed $\gamma_\delta$ is smaller than $c_\alpha$, we conclude that $\delta$ is a significant change point, indicating a change in network structure; otherwise, it is not a significant change point.  This procedure is repeated for each candidate change point.   

If the data are assumed to be independent and identically distributed, we perform a permutation inference procedure.  The permutation procedure is identical to the stationary bootstrap procedure above except we permute the data instead of resampling blocks of data.

\subsection{Choice of $K$}

In the spectral clustering step, the number of communities $K$ is a pre-specified parameter as the choice of $K$ in this framework remains an open problem.  However, recently progress has been made on this topic \citep[e.g.][]{FrancoEtal2014, ChenLei2014}.  Unfortunately, all the existing methods require extra computational resources.  In our method, the true number of communities is $K_o$ and we pre-specify an over-estimated $K$.  We will show in the numerical results that our method is robust with respect to $K$ and will perform uniformly well when $K$ is over-estimated.

We can understand this phenomenon from a dimension-reduction perspective.  Spectral clustering embeds a $p$-dimensional data set into a $K$-dimensional space.  When $K$ is under-estimated, important directions are missing.  One of the steps in our algorithm calculates the singular values between two modified dimension-reduced matrices.  Instead of using the matrices consisting of the principal components, we replace the rows by the centroid of the cluster it is in.  This, on one hand, makes the community structure more prominent; on the other hand, it further reduces the dimension from $p \times K$ to $K \times K$.  Ideally, for a $K_o$-community network, the principal component matrix is of rank $K_o$; therefore, if $K = K_o$, the two orthonormal matrices expand the basis of $\mathbb{R}^{K_o}$ space and the singular values are all 1.  

\section{Simulations}\label{sec:sims}

In this section we examine the performance of NCPD through various simulation settings.  For each setting, we perform 100 repetitions, provide a diagram to illustrate how the network structure changes over time and a quantified description of the distributional aspect.  To summarise the results in each setting, we plot the Gaussian kernel smoothed empirical density of the occurrences of the detected change points.  As we noted in Section~\ref{sec:selec-cri}, we require a certain number of points at the beginning of the time series to initiate the algorithm ($n_{\min}$).  During this period, we assume that the network structure is the same.  However, the time points close to the two ends of the time axis (data close to $n_{\min}$ and $T-n_{\min}$) tend to capture some network structural changes.  We call the occurrence of change points close to the points $n_{\min}$ and $T-n_{\min}$ the \emph{edge phenomenon}, and we delete change points that are significant but close to ends, and report the remaining false positive change points as \emph{modified false positives}.

To visualise the networks, we set a threshold for the correlation between two nodes.  If the absolute value of the correlation between two time series is larger than our pre-specified threshold (0.3 is used in the numerical results in this paper), we present the two corresponding nodes connected by an edge; otherwise, disconnected.  This is only for the sake of visualisation, while the weighted networks are decided without this threshold.

\subsection{Descriptions of the settings}\label{subsec:description}

\subsubsection{Network structure changes}

\begin{figure}[ht]
\centering
\makebox{\includegraphics[width = \textwidth]{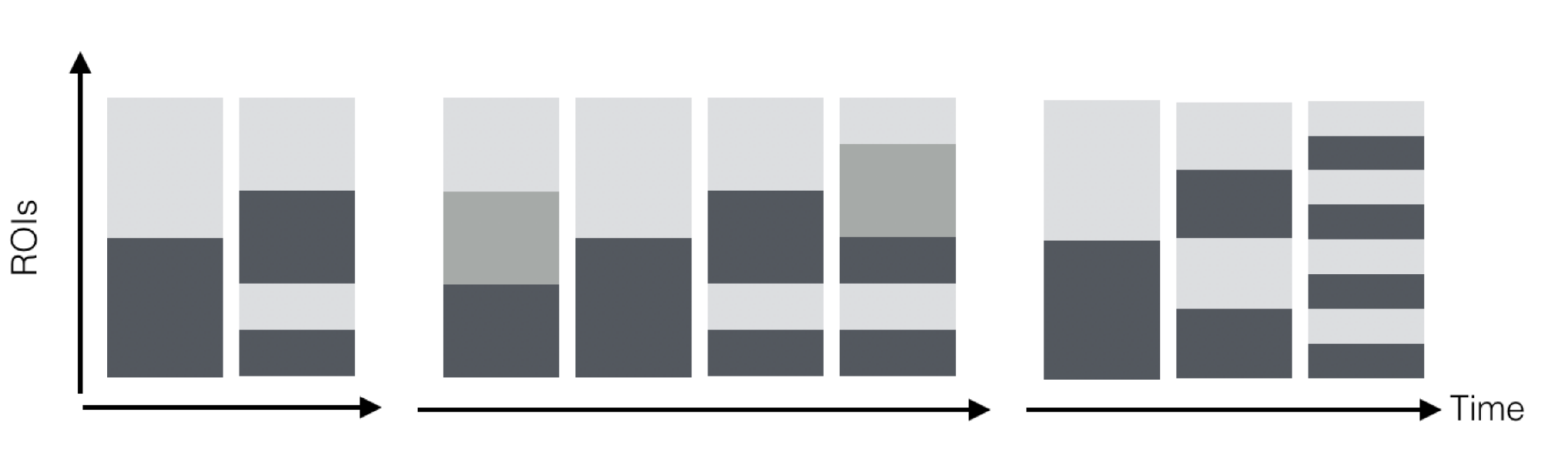}}
\caption{\label{design} Design of the simulations.  The left, middle and right panels are for settings 1, 2 and 3 respectively.  Within each setting, each rectangle represents a stationary process, and different colours within each rectangle represent different communities.  }
\end{figure}

In Figure~\ref{design}, we illustrate how the network structure changes in different settings.  The left, middle and right panels are for settings 1, 2 and 3, respectively.  
\begin{itemize}
\item In Setting 1, the change point occurs in the middle of the time series, with the true number of communities being $K_o = 2$ both before and after the change point.  At the change point, the vertices labels are randomly reshuffled.  
\item In Setting 2, there are three change points and they are located at the first, second and third quarters of the whole time line, respectively.  In the first time segment, the true number of communities $K_o = 3$, i.e., there are three clusters, one of which is equally merged into the other two clusters at the first change point.  Vertex labels are randomly shuffled at the second change point, while keeping $K_o = 2$.  The true number of communities $K_o$ returns to 3, by moving one third of each community into a third community.  
\item In Setting 3, two change points occur, with the true number of communities $K_o = 2$ remaining constant for the whole time course.  At each change point, half of the vertices in each community are moved to the other community. 
\end{itemize}

In terms of changing nature of the network structure, the easiest is Setting 1, where only one change point occurs and the community labels are reshuffled at the change point.  Setting 2 covers the situation where the true number of communities $K_o$ changes.  Setting 3 is the most challenging setup, where the structural changes only involve separating or merging existing communities.

\subsubsection{Distributional description}

\begin{itemize}
\item In Setting 1, $(p, T) = (400, 200)$ and the data are generated from the multivariate Gaussian distribution $\mathcal{N}(0, \Sigma)$, where
\begin{equation*}
\Sigma_{ij} = \begin{cases}
0.75, & \mbox{if } i\neq j \mbox{ and } i, j \mbox{ are in the same cluster};\\
 1, & \mbox{if } i = j;\\
0.20, & \mbox{otherwise}.
 \end{cases}
\end{equation*}

\item In Setting 2, $(p, T) = (600, 400)$ and the data are generated from the multivariate Gaussian distribution $\mathcal{N}(0, \Sigma)$, where
\begin{equation*}
\Sigma_{ij} = \begin{cases}
0.75, & \mbox{if } i\neq j \mbox{ and } i, j \mbox{ are in the same cluster};\\
 1, & \mbox{if } i = j;\\
 0.20^{|i-j|}, & \mbox{if } i, j \mbox{ are not in the same cluster}.\\ 
 \end{cases}
\end{equation*}

\item In Setting 3,  $(p, T) = (800, 600)$ and the data are generated from the same distribution as that in Setting 2.
\end{itemize}

\subsection{Results}

In this subsection, we present results in various formats.  Bearing in mind the fact that the detected change point may differ by a few time points from the true change point, we define those which are at most 10 time points away (either before or after) from the true change point as the \emph{true positives (TP)}.  We present the average number of TP across all 100 repetitions in each setting, along with the standard error (in brackets).  In addition, we present the frequency of the \emph{false positives (FP)}, as well as that of the modified FP, which excludes all the detected change points that are at most 10 time points away from $n_{\min}$, which is 50 in the simulations, and $T - n_{\min}$.   

\begin{figure}[htbp]
\centering
\makebox{
\includegraphics[width = 0.45\textwidth, height = 0.2\textheight]{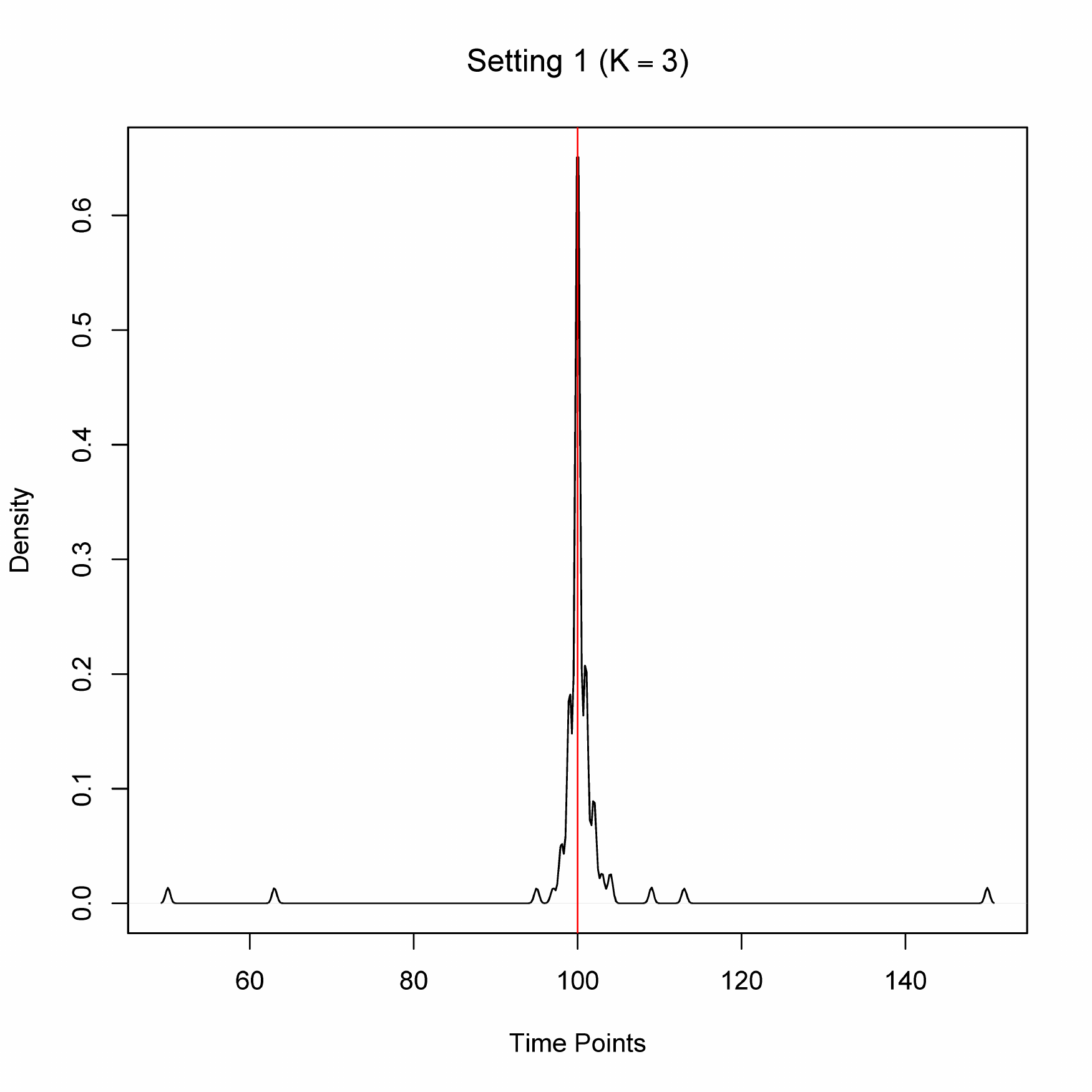}
\includegraphics[width = 0.45\textwidth, height = 0.2\textheight]{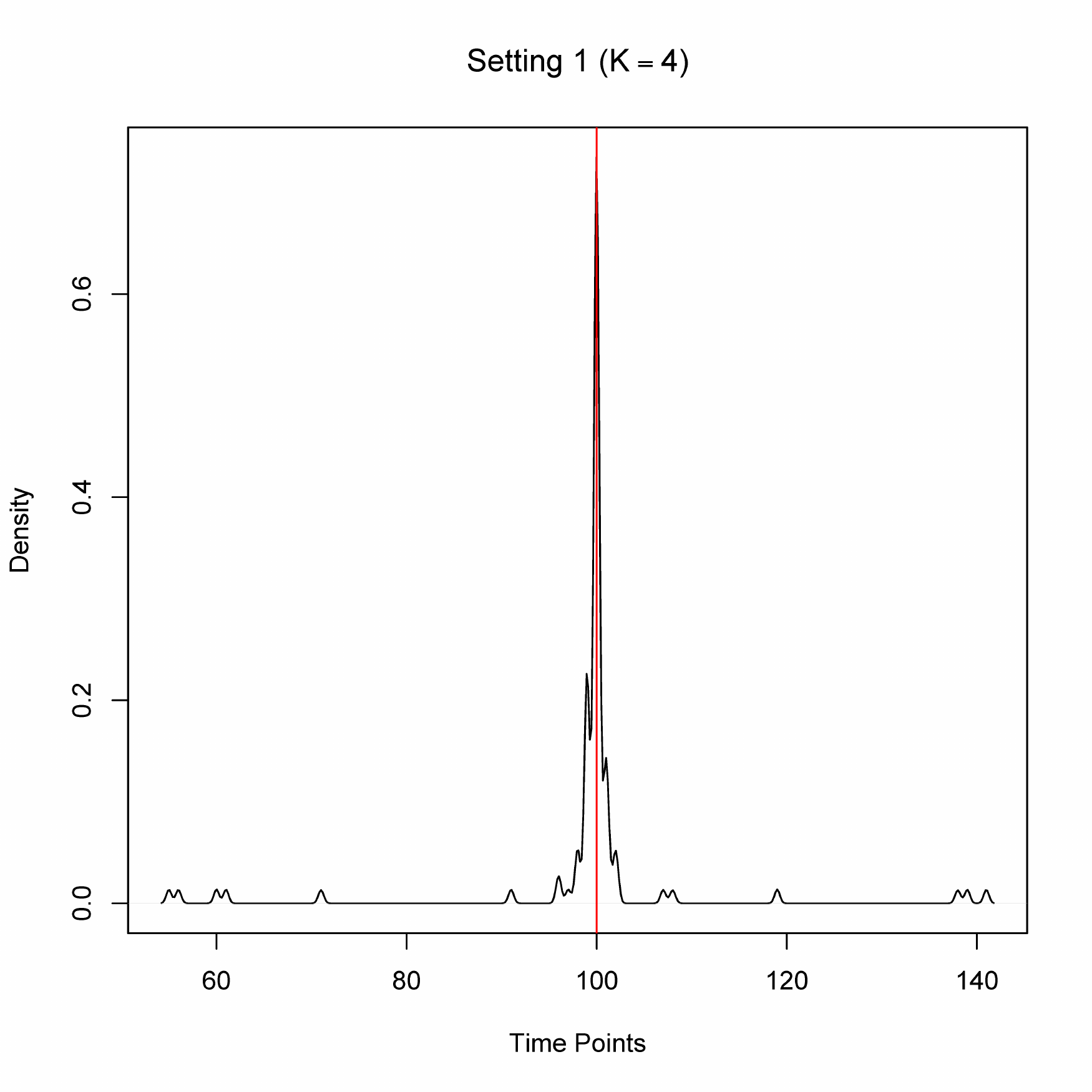}}
\makebox{
\includegraphics[width = 0.45\textwidth, height = 0.2\textheight]{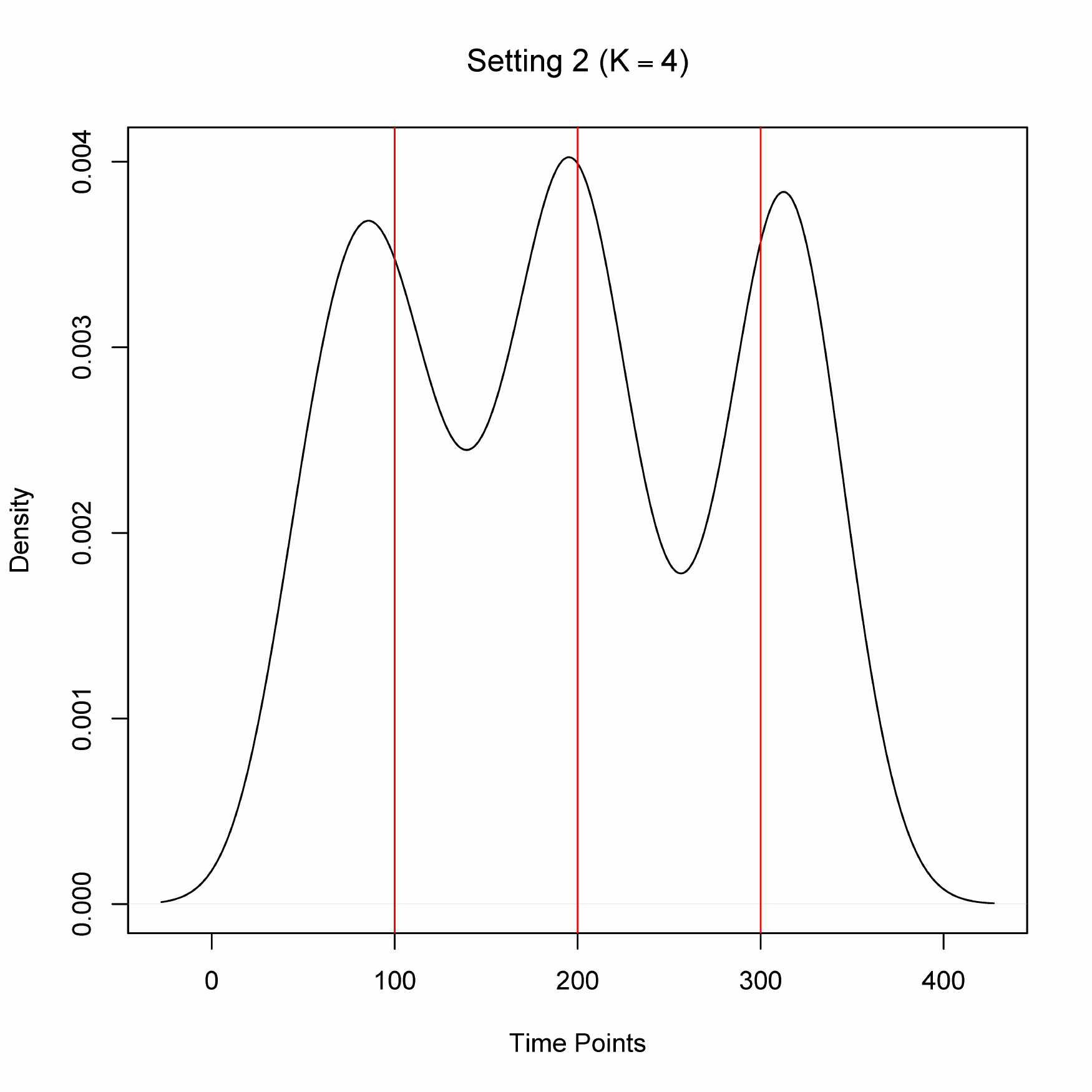}
\includegraphics[width = 0.45\textwidth, height = 0.2\textheight]{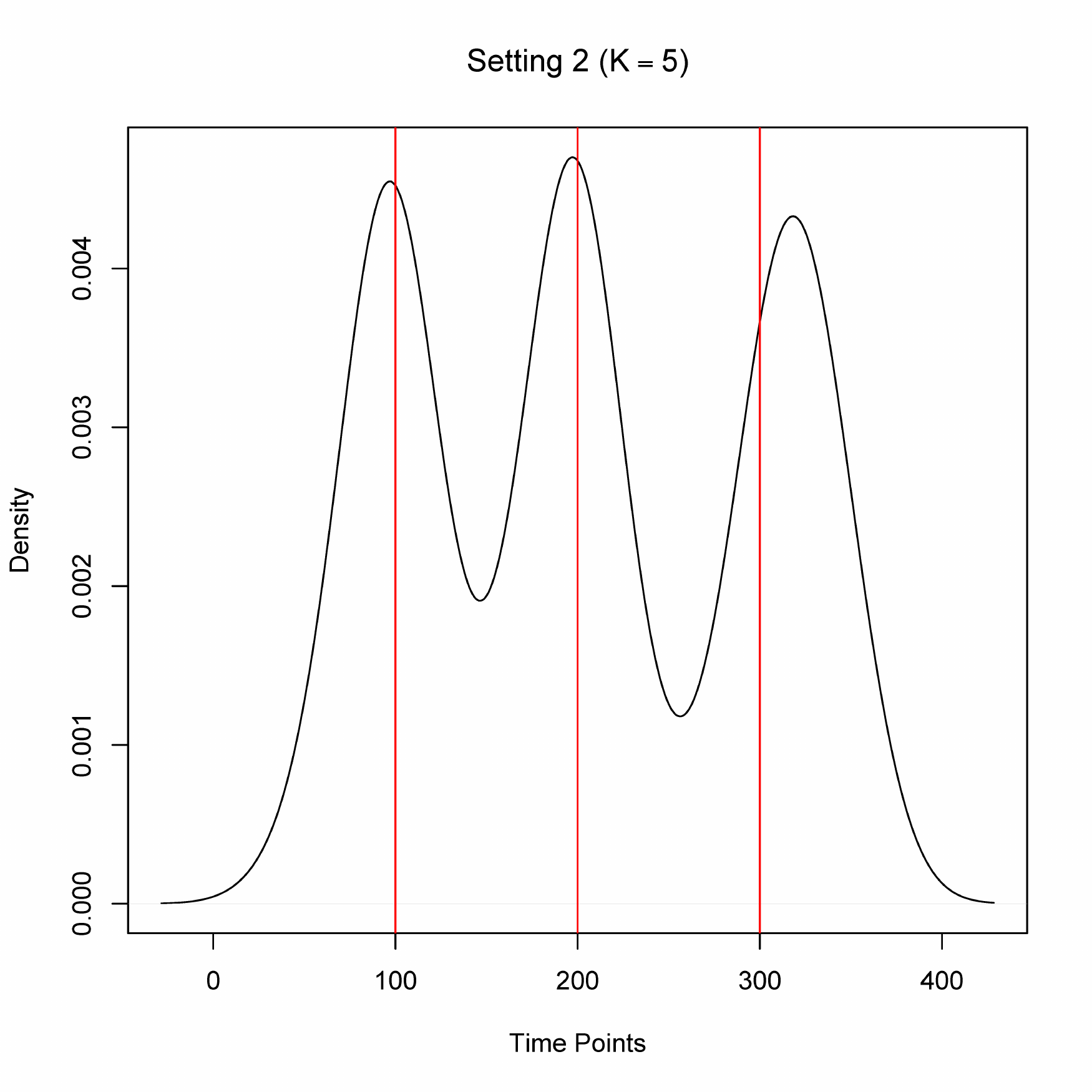}}
\makebox{
\includegraphics[width = 0.45\textwidth, height = 0.2\textheight]{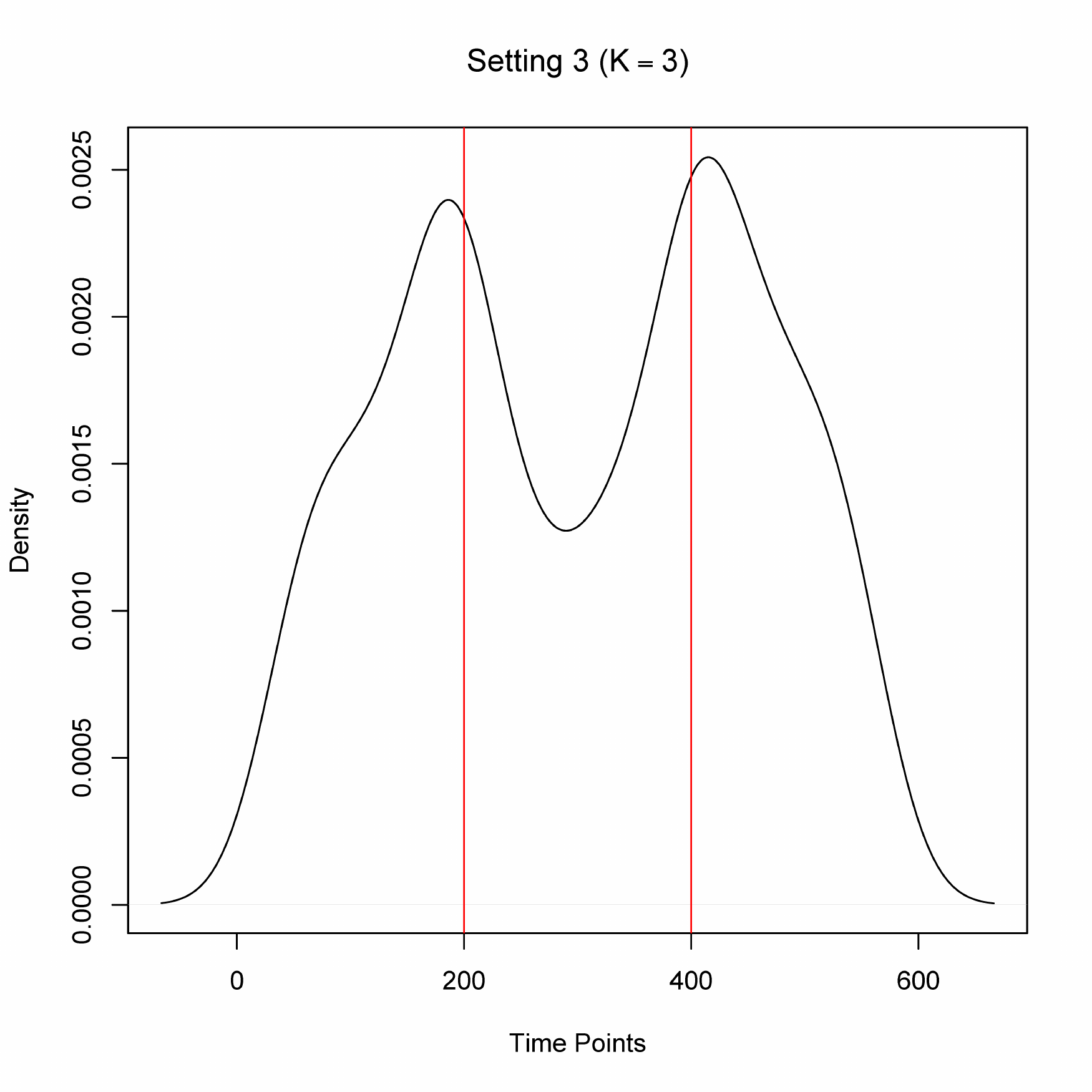}
\includegraphics[width = 0.45\textwidth, height = 0.2\textheight]{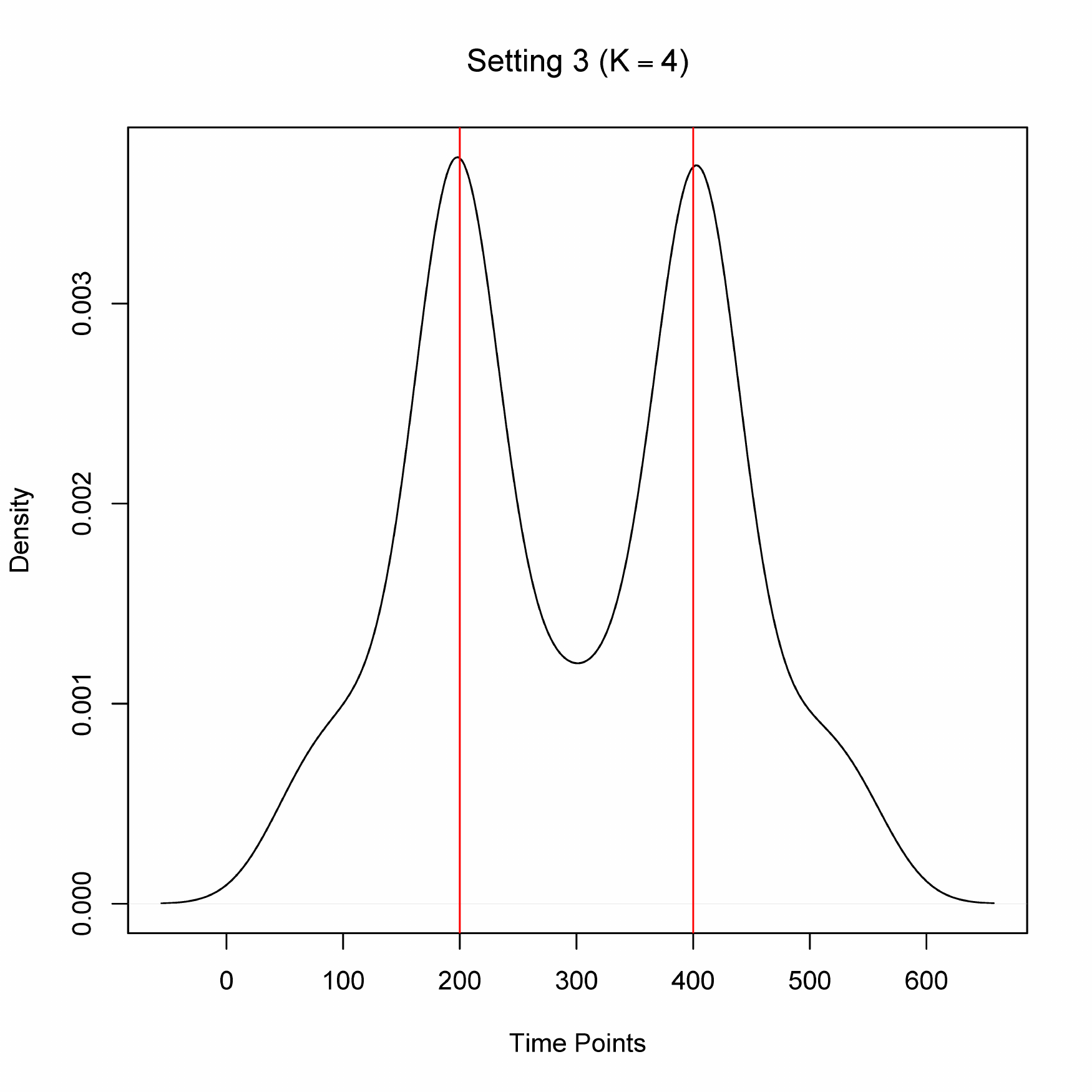}}
\caption{\label{fig:density} Gaussian kernel smoothed empirical density of the change point in all settings, with the red vertical lines indicating the true change points.}
\end{figure}

In Figure~\ref{fig:density}, we plot the Gaussian-kernel smoothed empirical densities of the change points for each of the simulation settings, with the red vertical lines indicating the true change points.  Notice that the true change point occurs at or near the peaks of the density curves in all the settings.  More quantitative results are collected in Table~\ref{tab:simulations}.  As we can see from the results, NCPD preforms well across all settings, with the true TPs all lying in the detected TP intervals.  Modified FP frequencies are significantly smaller than those of the FPs.  It is also worthwhile to point out that as long as $K$ is overestimated, NCPD performs robustly.

\begin{table}
\caption{\label{tab:simulations}Simulation results.  TP: true positives.  FP: false positives.  Freq: frequencies.  mod: modified.}
\fbox{%
\centering
{\small
\begin{tabular}{*{9}{c}}
Setting & $K_o$ & $K$ & \multicolumn{3}{c}{detected TP(s)} & TP Freq. &  FP Freq. & mod. FP Freq. \\
\hline 
1 & 2 & 3 & 100.23 (1.54) & &  & 0.99 &  0.04 & 0.00 \\
1 & 2 & 4 & 99.89 (1.75)  & & &  0.97 & 0.06 & 0.00\\ 
2 & 2 \& 3 & 4 & 97.52 (4.21) & 198.86 (4.18) & 302.41 (4.18) & 1.40 & 2.35 & 2.05\\
2 & 2 \& 3 & 5 & 102.53 (4.81) & 199.05 (4.03) & 299.89 (3.22) & 1.35 & 1.55 & 1.43\\
3 & 2 & 3 & 197.40 (3.77) & 401.64 (4.02) &  & 1.15 & 4.24 & 3.05\\
3 & 2 & 4 & 198.45 (3.20) & 401.58 (3.11) & & 1.73 & 1.49 & 0.50\\
\end{tabular}} }
\end{table}

\begin{figure}
		\centering
				\makebox{
						\includegraphics[width = \textwidth]{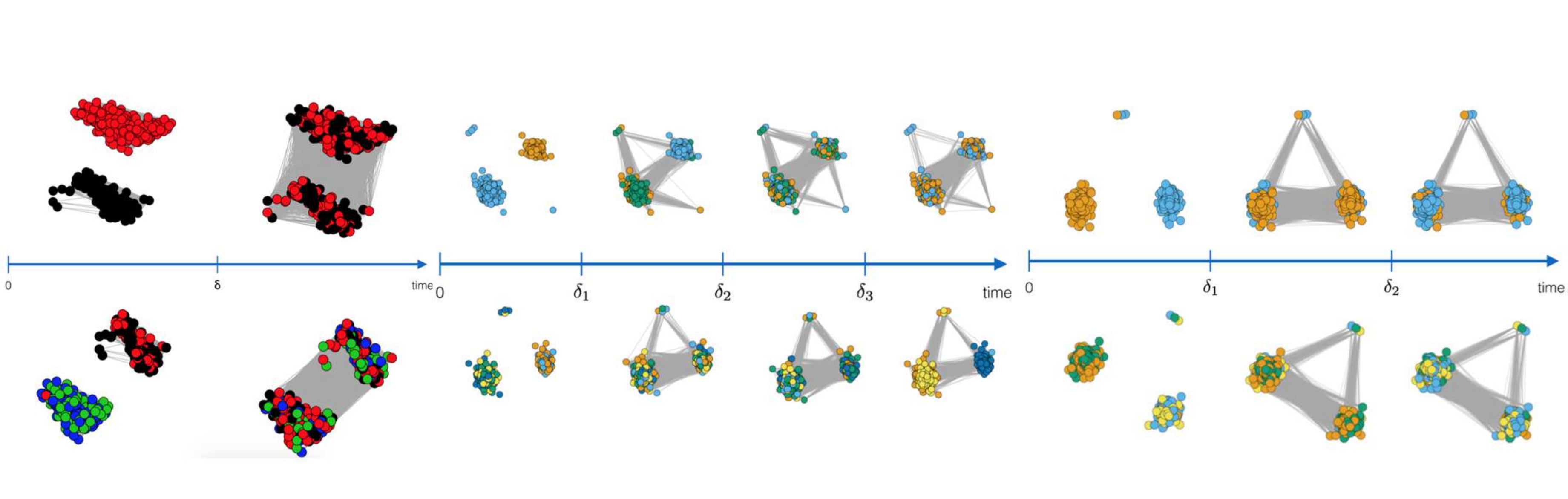}}
				\caption{\label{fig:networks} Network plots of Settings 1, 2 and 3.   In the lower panel, the networks before and after the detected change point by using the pre-specificied number of communities are plotted.  In the upper panels, the true change point and the networks using the true number of communities $K_o$ in each partition are plotted.}
\end{figure}

In addition, we present the network graphs.  In Figure~\ref{fig:networks}, we pick one realisation for each setting.  The left, middle and right panels are representatives of settings 1, 2 and 3, respectively.  In the lower panel, we plot the networks before and after the detected change point.  The specific number of communities in the lower panel are $K = 4, 5$ and 4, respectively, i.e. $K$ different colours indicating $K$ different communities.  Let us take the left panel, which represents Setting 1 as an example.  It is obvious that in the lower-left panel, which represents the network before the change point, the blue and green nodes belong to the same group, and most of the black and red nodes belong to the other.  There are no connections between these two groups.  After the change point, the network is visualised in the lower-right panel, with the same layout of the vertices.  In this case, these two groups are well-connected and the four different colours are mixed between the two groups.  In the upper panels, we present the true change point and plot the network using the true number of communities $K_o$ in each part.  In the left panel, we can see that the red and black nodes are well-separated in the network prior to the change point, while they are mixed between communities in the network after the change point. 

\section{Resting-state fMRI data}\label{sec:rsfmridata}
We apply NCPD to a resting-state fMRI data set, as described in \cite{habeck}.  Participants ($n=45$) are instructed to rest in the scanner for 9.5 minutes, with the instruction to keep their eyes open for the duration of the scan.  We apply the Anatomical Automatic Labeling \citep{tzourio} atlas to the adjusted voxel-wise time series and produce time series for 116 Regions of Interest (ROIs) for each subject by averaging the voxel time series within the ROIs.  In total, each time series contained 285 time points (9.5 minutes with TR = 2). 

Table~\ref{tab:real} shows the significant change point locations for all 45 subjects.  Every subject except one (subject 16) has at least two significant change points in their community network structure with the maximum number of change points being 4.  The table indicates that not only does the number of community network change points differ across subjects, but the location of the change points is also variable.  In addition, some subjects remain in states for long periods while others transition more quickly. 
Hence, the method has a major advantage over moving-window type methods as we do not have to choose the window length, which can have significant consequences on the estimated FC. 

We used 1,000 stationary bootstrap resamples of the data for inference on the cosine values of the principal angles at each candidate change point with an average block size of 57 (or 20\% of the data set) and the minimum distance between change points, $n_{\min}=50$.  Previous work \citep{allen} found the existence of 7 resting-state networks; hence, we specified $K=7$ clusters or communities in our algorithm.   The time taken to run the algorithm on each subject ($T=285$, $p=116$) using a Intel(R) Core(TM) i5-3210M 2.50GHz CPU was on average 132s.

\begin{table}
\caption{\label{tab:real}Resting-state fMRI data}
\fbox{
\centering
\begin{tabular}{*{4}{c}}
 	Subject & Significant CP location & Subject & Significant CP location \\ \hline
			1       & 57, 133, 208            & 24      & 79, 167, 230       \\ 
			2       & 82, 170, 233            & 25      & 51, 118, 209       \\ 
			3       & 54, 114, 167, 220       & 26      & 53, 108, 166, 233  \\ 
			4       & 86, 159, 211            & 27      & 53, 107, 181, 233  \\ 
			5       & 54, 104, 169, 233       & 28      & 59, 118, 168, 218  \\ 
			6       & 80, 161, 231            & 29      & 101, 152, 203      \\ 
			7       & 59, 131, 232            & 30      & 55, 108, 186       \\ 
			8       & 56, 147, 216            & 31      & 96, 149, 203       \\ 
			9       & 84, 150, 229            & 32      & 58, 108, 208       \\ 
		  10      & 52, 118, 172, 233       & 33      & 63, 133, 183, 234  \\
		  11      & 69, 119, 177, 231       & 34      & 68, 144, 227       \\
			12      & 70, 123, 183, 233       & 35      & 53, 123, 221       \\ 
			13      & 54, 108, 161, 225       & 36      & 51, 107, 157, 233  \\ 
			14      & 52, 116, 166, 232       & 37      & 68, 118, 176, 229  \\ 
			15      & 93, 192                 & 38      & 62, 119, 173, 235  \\
			16      &                         & 39      & 69, 120, 189       \\ 
			17      & 53, 117, 169, 231       & 40      & 82, 137, 215       \\ 
			18      & 50, 100, 155, 230       & 41      & 70, 122, 180, 234  \\ 
			19      & 50, 132, 184, 234       & 42      & 64, 125, 191       \\ 
			20      & 51, 113, 168, 224       & 43      & 52, 131, 213       \\ 
			21      & 54, 109, 165, 230       & 44      & 58, 128, 214       \\ 
			22      & 50, 118, 177, 230       & 45      & 88, 155, 215       \\ 
			23      & 77, 136, 230            &         &                    \\ 
     \end{tabular}}
\end{table}

\begin{figure}[h]
		\centering
				\makebox{\includegraphics[scale=0.4]{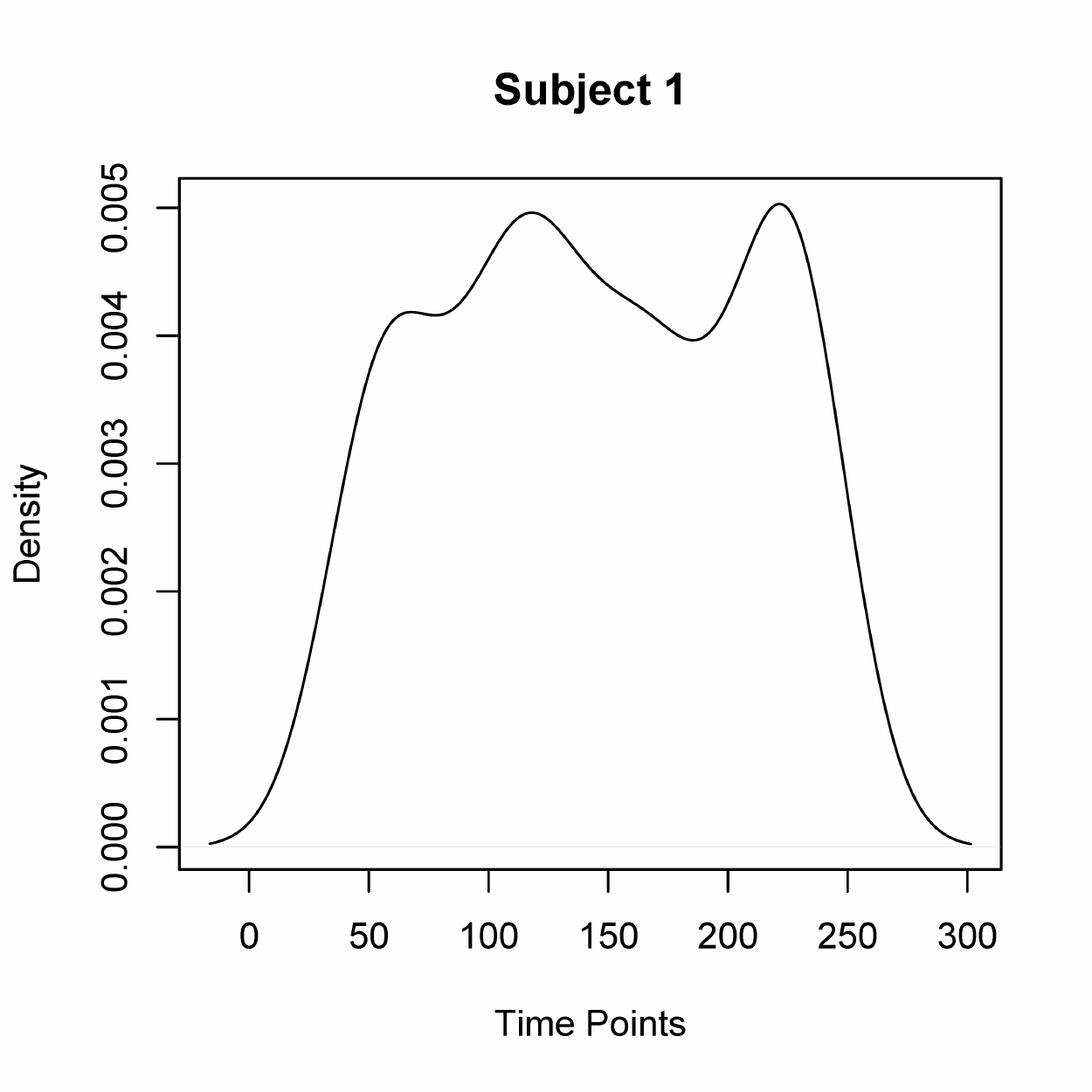}}
				\makebox{\includegraphics[scale=0.4]{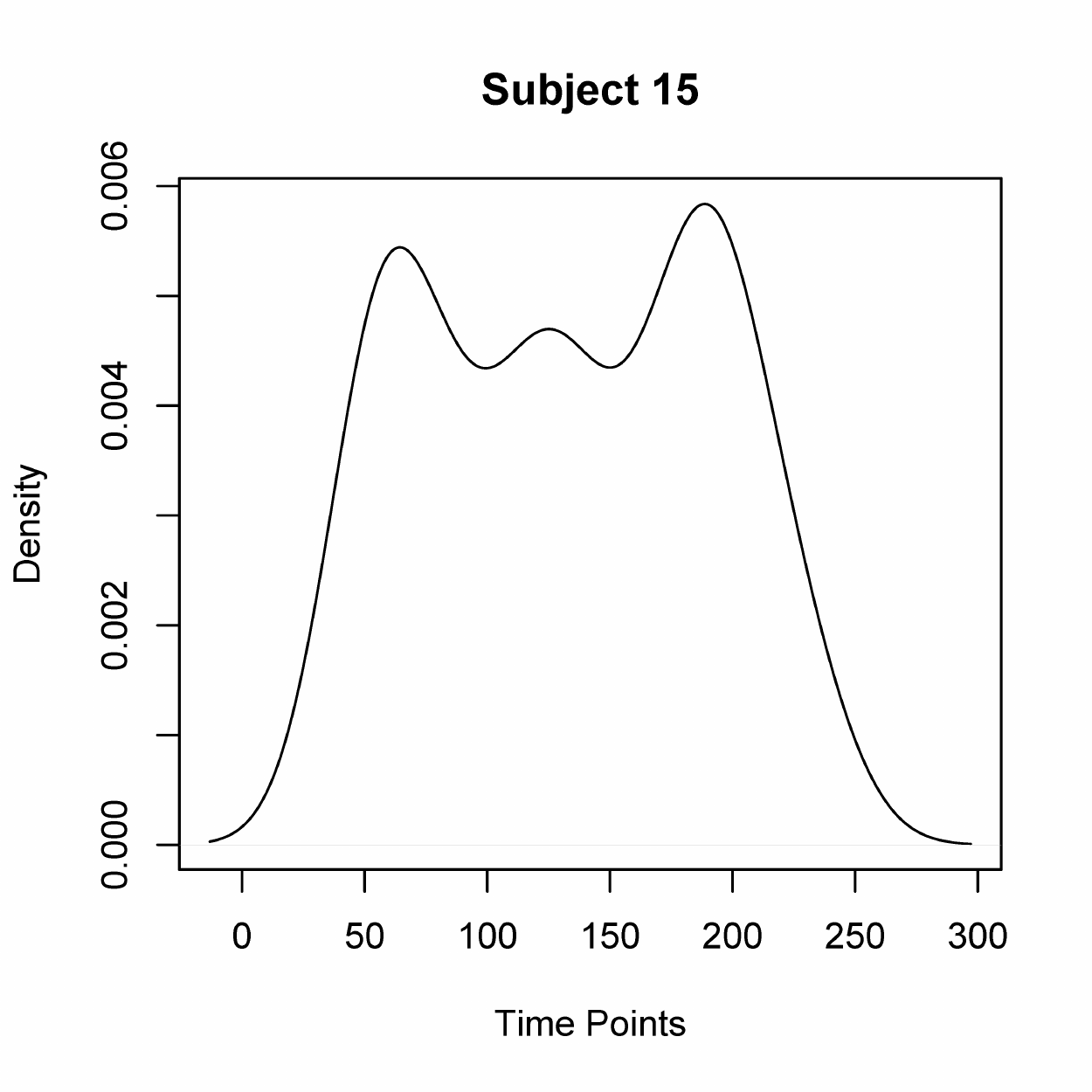}}
				\caption{\label{fig:caplets} Density plots of the detected change points in sub-sampling data from Subjects 1 and 15 in the resting-state fMRI data set.}
\end{figure}

In Figure~\ref{fig:caplets}, we mimic the idea in the simulation study to plot the empirical density of the detected change points.  In order to get repeated samplings, every time we delete 10\% of the data sequentially and therefore have 10 repetitions for each data set.  We can see that these is a bump around time point 100 in both subjects (taking the edge phenomenon into consideration), which is a change point we should be cautious about.

\begin{figure}[ht]
				\makebox{\includegraphics[scale=0.34]{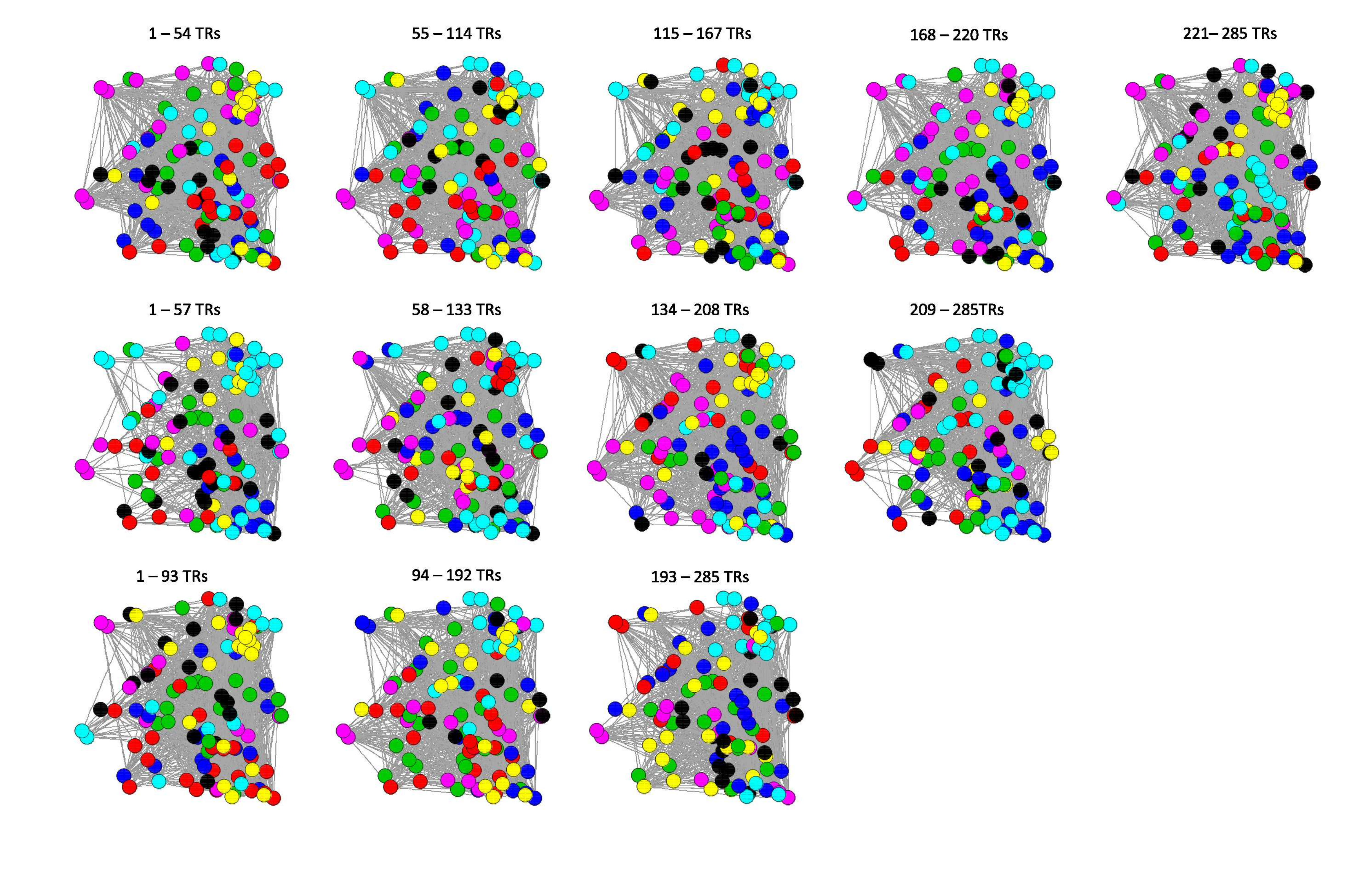}}
				\caption{\label{fig:netplots} The corresponding community network graphs for the resting-state fMRI study. The graphs on the first, second and third rows represent the graphs for subjects 3, 1, and 15, respectively. The significant change points can be found in Table~\ref{tab:real}.}
\end{figure}

Figure~\ref{fig:netplots} shows the estimated community network graphs for data between each pair of significant change points shown in Table~\ref{tab:real}.  The graphs on the first, second and third rows represent the graphs for subjects 3, 1, and 15, respectively.  The colour of the node represents which community the ROI time series belongs to. 
 
\subsection{Cross subject comparisons}

As the data are from a resting-state study when mental activity is unconstrained, we do not expect the community network structure for each subject to match along the same partitions in all cases, that is, we do not anticipate that subject 3's community network in their first partition to be similar to subject 15's community network in their first partition.  However, we do foresee similarities across the subjects' partition plots.  In particular, we assume that subjects will enter some common stable functional `states' or community network patterns. 

\begin{figure}[ht]
				\makebox{\includegraphics[scale=0.36]{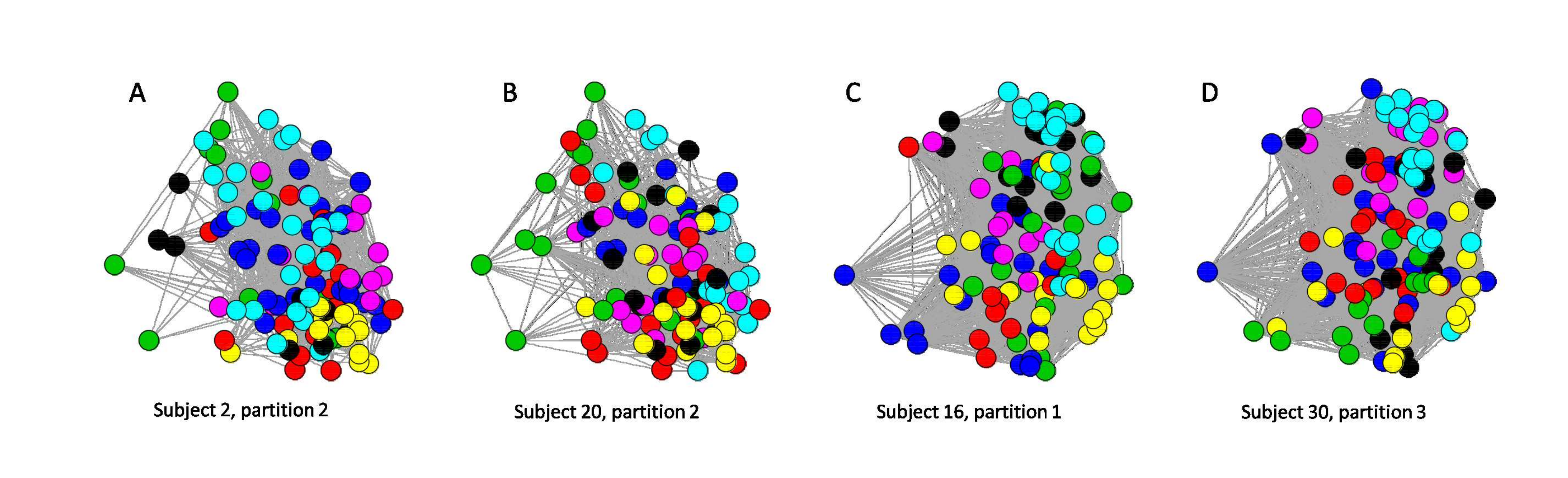}}
							\[
				\begin{blockarray}{cccccccc}
					& S2(1) & S2(2) & S16(1) & S20(1) & S20(2) & S30(2) & S30(3)  \\
					\begin{block}{c(ccccccc)}
								S2(1)  &      & 1.85 & 1.70 & 1.58 & 1.59          & 1.12 & 1.49 \\
								S2(2)  &      &      & 2.16 & 1.80 & \textbf{2.19} & 1.36 & 1.96 \\
								S16(1) &      &      &      & 1.43 & 1.35          & 1.38 & \textbf{2.15} \\
								S20(1) &      &      &      &      & 1.65          & 1.35 & 1.68 \\
								S20(2) &      &      &      &      &               & 1.48 & 1.97 \\
								S30(2) &      &      &      &      &               &      & 1.50 \\
								S30(3) &      &      &      &      &               &      &      \\
				\end{block}
		\end{blockarray}
 \]
				\caption{\label{fig:partnetplots} Figures A, B, C and D are examples of the community network patterns for subject partitions.  Figure A and B, and Figure C and D appear to have similar network patterns.  In the matrix, we calculate the criterion values for the similarity of networks across subjects.  S2(1) represents subject 2, partition 1. For example, the criterion value for the two networks in Figure A and B is located in element (2,5) of the matrix.  The higher the values, the more similar the networks.  }
\end{figure}

In Section \ref{sec:modelal}, we use the singular values of the product matrix, $U_{\mathrm{L}}^{\top}U_{\mathrm{R}}$, consisting of the matrices (networks) before and after a certain time point, to detect the change point.  The rationale for this is that the criterion value represents the similarity of the networks.  In the same spirit, when estimating the similarity of the networks across different subjects, we can use the singular values of $U_{i,j}^{\top}U_{k,l}$ where $U_{i,j}$ is the transformed network from subject $i$, partition $j$ and $U_{k,l}$ is the transformed network from subject $k$, partition $l$.  In Figure~\ref{fig:partnetplots}, we calculate the criterion values, $U_{i,j}^{\top}U_{k,l}$, between a small number of cross-subject functional state pairs.  The higher the criterion values in the matrix, the more similar the networks within or across subjects.  For example, the criterion value for the two networks for subject 2, partition 2 and subject 20, partition 2 (Figures A and B) is located in element (2,5) of the matrix and represents two of the most similar across subject state-pairs. 

To show (graphical) evidence of common functional states across subjects, we also plot in Figure~\ref{fig:partnetplots} the community network structure for subject 2 (partition 2), subject 20 (partition 2), subject 16 (partition 1) and subject 30 (partition 3).   We compare the similarity of Figures~\ref{fig:partnetplots} A and B and Figures~\ref{fig:partnetplots} C and D.  In particular, we compare the colour patterns of the graphs.  For example, in Figure~\ref{fig:partnetplots}, the green, yellow and red nodes in A and the green,  yellow and red nodes in B are very similar. There are also similar crossovers between the aqua, pink and blue nodes.  Hence, we conclude that the subjects enter a similar cognitive or functional state.  There is also similarity across Figures~\ref{fig:partnetplots} C and D; here, the aqua nodes in C are very similar to the aqua nodes in D.  This is also true for the yellow and blue nodes in both C and D.  There are many more examples of this in the data set.  

The resting-state data set in this paper contains regional time series from several networks including the Default Mode Network, Dorsal Attentional Network, Executive Control Network, Senorimotor Network, Visual Network, Auditory Network and Salience Network.   However, the common cognitive states or structured patterns found across subjects do not directly link up with these networks, therefore, providing us with observed states that relate differently to previous findings.  Moreover, the common structured patterns or functional states found by NCPD include communities between regions not attributed to the list of networks and communities with regions from a few of the networks.  In particular, some of the common functional states found show that some communities have strong synchrony across the different networks and weak synchrony with other regions from the same network.  Hence, the features found are significant and meaningful given the fact that this is the first study to consider over a hundred fMRI resting-state time series.  However, we remain cautious because resting-state fMRI is unconstrained in nature and the functional roles of dynamics and their relationship to subjects' cognitive state remains unknown.  We believe that further investigations into the specificity and consistency of the fMRI functional states or features will be beneficial and that work to elucidate spatiotemporal dynamics associated with spontaneous cognition and behavioural transitions is very important. We hope that NCPD will add to this endeavour.

\section{Discussion} \label{sec:discussion}
In this paper, we develop a new approach, NCPD,  for analysing and modelling multivariate time series from an fMRI study which consists of realisations of complex and dynamic brain processes.  The method adds to the literature by improving understanding of the brain processes measured using fMRI.  NCPD is, to the best of our knowledge, the first paper to consider estimating change points for time evolving community network structure in a multivariate time series context.  

NCPD is an innovative approach for finding psychological states or changes in FC in both task-based and resting-state brain imaging studies.  There are several novel aspects of NCPD. Firstly, it allows for estimation of dynamic functional connectivity in a high-dimensional multivariate time series setting, in particular, in situations where the number of brain regions is greater than the number of time points in the experimental time course ($p>T$).  Hence, it can consider the dynamics of the whole brain or a very large number of ROI or voxel time series, thereby providing deeper insights into the large-scale functional architecture of the brain and the complex processes within.  Secondly,  it is not restricted by the situations that commonly occur in change point settings, such as the at most one change (AMOC) setting and the epidemic setting (two change points, where the process reverts back to the original regime after the second change point). Indeed, NCPD is flexible as there is no \textit{a priori} assumption on the number of changes and where the changes occur.  Finally, NCPD is, to the best of our knowledge, the first piece of work to consider estimating change points of time evolving community network structure in a multivariate time series context.  We introduced a novel metric to find the candidate change points, i.e. the singular values of the product matrices formed by the before and after change point networks.  However, NCPD is restricted by the minimum distance between change points (the $\delta$ parameter in the algorithm).  

It has been shown that neurological disorders disrupt the functional connectivity pattern or structural properties of the brain \citep{greicius,menon}.  Future work entails applying NCPD to subjects with brain disorders such as depression, Alzheimer's disease and schizophrenia and to control subjects who have been matched using behavioural data. NCPD may lead to the robust identification of cognitive states at rest for both controls and subjects with these disorders.  It is hoped that the large-scale temporal features resulting from the accurate description of functional connectivity from our novel method will lead to better diagnostic and prognostic indicators of the brain disorders.  More specifically, by comparing the change points and the community network structures of functional connectivity of healthy controls to patients with these disorders, we may be able understand the key differences in functional brain processes.  In particular, NCPD allows us to find common cognitive states that recur in time, across subjects, and across groups in a study.

While NCPD is applied to resting-state fMRI data in this work, it could seamlessly be applied to an Electroencephalography (EEG) or Magnetoencephalography (MEG) data set.  Moreover, NCPD pertains to a general setting and can also be used in a variety of situations where one wishes to study the evolution of a high dimensional network over time. 

NCPD appears to have a large computational cost with the binary segmentation of the data and the stationary bootstrap procedure for inference on the candidate change points. However, the resting-state fMRI data set shows how fast the algorithm is.  Both the binary segmentation and the stationary bootstrap procedures use parallel computing and on a dual core processor the average time to run the algorithm on one subject ($T=285$, $p=116$) was 132s.  Obviously, with access to more cores this time will decrease significantly.  The code for the algorithm can already be downloaded from  http://www.statslab.cam.ac.uk/$\sim$yy366/.

\section{Acknowledgement}

We sincerely thank the Editor, the Guest Editor and two referees for their constructive comments.  Funding for Ivor Cribben was provided by the Pearson Faculty Fellowship (Alberta School of Business) and a Alberta Health Services (AHS) Grant.  Yi Yu is supported by the Professor Richard Samworth's Engineering and Physical Sciences Research Council Early Career Fellowship EP/J017213/1.

\bibliographystyle{plain}
\bibliography{jrssc-final}

\end{document}